\documentclass[twocolumn,showpacs,preprintnumbers,amsmath,amssymb]{revtex4}

\begin{document}
\title
{
\bf Designing molecules to bypass the singlet-triplet bottleneck
in the electroluminescence of organic light-emitting-diode materials.
}

\author
{
 M.W.C. Dharma-wardana}
\email{chandre@nrcphy1.phy.nrc.ca}
\affiliation{Institute for Microstructural Sciences, National Research Council, Ottawa,Canada. K1A 0R6}
\author{ Marek Z. Zgierski}
\affiliation{ Stacie Institute of Molecular Sciences,National Research Council, Ottawa,Canada. K1A 0R6}

\date{27 January 2002}
\begin{abstract}
Electroluminescence in organic light emitting diode (OLED) materials
occurs via the recombination of excitonic electrons-hole pairs.
Only the singlet excitons of commonly used  OLED materials,
e.g., Aluminum trihydroxyquinoline (AlQ$_3$), decay radiatively,
limiting the external quantum efficiency to a maximum 25\%.
Thus 75\% of the energy is lost due to the triplet bottleneck
for radiative recombination.
We consider molecules derived from AlQ$_3$ which bypass the
triplet bottleneck by designing structures which contain
strong spin-orbit coupling. As a first stage of this work, we calculate the
groundstate energies and vertical excitation energies of 
Al-arsenoquinolines and Al-boroarsenoquinolines.
It is found that the substitution of N by As leads to
very favourable results, while the boron substitution leads
to no advantage.
\end{abstract}
\maketitle



\section{Introduction}
Increased interest in the use of organic materials\cite{levine}
 for a variety of optical and electronic applications, including the
fabrication of electroluminescent devices, field-effect transistors,
lasers etc., 
has followed the successful demonstration of
devices with useful
 lifetimes and performance.\cite{tang87,jap95,at&t95,friend,brown}
 The new
advances depend on  a better choice of organic materials, and on new
technologies borrowed from recent 
developments in  semiconductor technology,
thin films, surface science and materials preparation. 
The luminescence of organic light emitting diodes (OLED)
 proceeds by the injection of electrons and holes
which form excitons and then  decay radiatively to the
ground state.\cite{nilar}
 Although better design of the diode structure can
improve the quantum efficiency of the device considerably,
selection rules for optical transitions in these organic materials
limit the radiative channel to singlet excitons. Hence the three-times
more abundant triplet excitons decay by {\it nonradiative} channels and
are wasted. This raises the possibility of an enormous gain in efficiency
if the triplet excitons could also be harvested for light emission.
 Baldo et al. attempted to
tap the triplet excitons by energy transfer from the host organic to a
porphine-platinum flourescent dye.\cite{baldo}

The objective of this work is to design the {\it host material itself}
 so that the
singlet-triplet selection rules become irrelevant. This is done using
heavy atoms where spin-orbit interactions ($L-S$ coupling) become important,
so that $S^z$ is no longer a good quantum number. We note in passing that
the spin selection rules cannot be circumvented by applying magnetic
fields, or incorporating local magnetic impurities since the spin
remains a good quantum number even in the presence of such fields.
It is necessary to bring the electrons into atomic sites where
the spin-orbit coupling intervenes to break the spin selection rules.
Since metal chelates
(e.g., Al, In, Sc) of 8-hydroxyquinoline have provided some of the
most successful OLED materials, we will consider molecules based
on modifications of these standard materials. 

\section{New molecules}
 Since  AlQ$_3$, i.e., 
Al(8-hydorxyquinoline)$_3$, has been a very successful host material (fig.~1)
in OLEDS,
we consider two structures derived from AlQ$_3$ where the quinoline part
is modified.  These are shown in Figs.~2 and 3, and are obtained by
first  replacing the N atom by an As atom ( an arsenoquinoline, denoted by
Al(AsQ)$_3$ ), and then replacing the carbon atom {\it para} to
the As atom by a boron atom, cf., fig.~3, to give Al(AsBQ)$_3$.
We proceed to investigate if such molecules are stable, and what
luminescence properties could be expected from them, using sophisticated
state of the art quantum chemical calculations. These molecules were
chosen for this investigation following detailed studies of AlQ$_3$
itself. For example, on examining the electron densities of the ground state
and the excitonic states of AlQ$_3$, it is found that the Al ion itself
 does not play a strong role in determining the exciton density.
 That is, replacing the
Al by a heavy chelating atom where the $L-S$ coupling is strong would not
help the luminescence. This is also experimentally known to be true, e.g.,
in the Scandium complex. On the other hand, the N atom in the quinoline ring
participates decisively in the relevant excitonic states. Thus replacing the
N atom by As is appealing. Of course, the spin-orbit interaction is
even stronger in Sb, but replacing N by Sb would be too strong a
chemical modification and the synthesis of the corresponding molecule may be
even more difficult. hence here we examine the case of replacing N by As.
Preliminary calculations showed that the N$\to$As
substitution had the effect of shifting the luminescence to the red, while the
shorter wavelength of the AlQ$_3$ is more attractive for device applications.
It was envisaged that replacing the {\it p-}C atom, i.e, the C atom in the
{\it para}-position to the N-atom, by boron would shift
 the spectrum back to the shorter wavelength regime. In this communication
we examine these questions using sophisticated quantum calculations. We find
that the N$\to$As leads to a strong improvement in the spectrum, without
too much of an adverse red shift , while the
further substitution of the {\it p-}C atom by B is not useful.\\

\section{Results}
The ground states of the metal chelates were calculated using the Gausssian-98
code at the B3LYP/3-21G* level.\cite{gaus,basis}. Both Al(AsQ)$_3$ and
Al(AsBQ)$_3$ were found to be stable. Hence it should be possible to
achieve a synthesis of these materials. The vertical excitation spectra
were calculated at the TDB3LYP/3-21G* level.\cite{basis1}. At this
point we have {\it not}
included spin-orbit coupling as yet. This is because the chelate structures,
containing three 8-hydroxyquinoline groups,
are too large for a direct implementation of relativistic effects. This would
be undertaken as the next stage of this investigation. In anticipation of
those results we have simply replaced the triplet oscillator strengths which
are zero  by
an analogous singlet oscillator strength and plotted the excitation
spectra ( see also table I).
Some justification for the assumption that the singlet $\to$ groundstate
oscillator strength and that for the triplet $\to$ ground 
become very similar in the
presence of the spin-orbit interaction in As is given in the
appendix.  The resulting enhanced triplet+singlet spectra as
well as the standard (singlet only) spectra are shown in Fig.~4

The excitations marked with an asterisk in Table I, for AlQ$_3$ and
Al(AsQ)$_3$ are transitions which involve the H (i.e., HOMO) and L (i.e, LUMO) 
orbitals in a major way. Thus in AlQ$_3$, the first singlet, i.e., S1,
 is approximately 0.98H $\to$ L + other configurations.
while the T4 is approximately 0.96H $\to$ L + other configurations. 
In the As substituted
form Al(AsQ)$_3$ we have T4 being 0.9H $\to$ L + other configurations, while the
S1 is higher in energy and is an admixture with
 0.94H $\to$ L + other configurations.
Hence these transitions may be identified with the HOMO-LUMO gaps used in
simplified theoretical schemes. Such simplified schemes are of course
modified when CI is included, where in various excited states (based on the
ground state determinant) get included. The deficiencies of the simpler
schemes become
 quite evident when we
go to the boron substituted form Al(AsBQ)$_3$. Here it is not possible to
identify a transition where the initial  state is mostly
 HOMO (i.e., $>$ 90\%).
In fact, T5, at 2.105 eV has 0.9H but the final orbital is L+4 admixed with
other configurations. The H $\to$ L transitions are at T8 with 0.76H in the initial
orbital, and S4 with 0.79H in the initial orbital. This shows that the B substitution
effectively introduces transitions in the nominal H $\to$ L gap of simplified
theories. Four additional transitions of Al(AsBQ)$_3$, which might be labeled
T10, S7, S8, and S9 in the notation of Table I are at 2.908, 2.913, 3.057,
and 3.103 eV and complete the same energy window as in the other two compounds.

Comparison of the calculate Al(Q)$_3$ spectra with experiment
 is not shown here, since such a comparison was carried out in a
previous publication.\cite{nilar} Such comparison require
 taking due account
 of solvent effects, or aggregation effects etc, and as well as considerations
of non-radiative pathways, since the experimental spectral intensities
are affected by many such factors. The objective of this study is to
examine the  Al(Q)$_3$,  Al(AsQ)$_3$, and  Al(AsBQ)$_3$ spectra calculated
under the {\it same } conditions, and in the same energy window.

 These results show the  Al(AsQ)$_3$  spectrum in the same energy
 window as  Al(Q)$_3$; however, its spectral features are more intense.
The main intensity is red shifted by 0.3 eV, confirming the conclusions
based on more elementary considerations. However, there are now
transitions in the blue as well. The substitution of the
{\it p-}carbon by boron, i.e, Al(AsBQ)$_3$, shows that the stronger 
absorption lines are somewhat blue shifted. The center of gravity of the
spectral intensity is 
 slightly blue shifted, although many peaks in the middle of the energy
 range are red shifted. The substitution of a nominally trivalent 
boron atom for a nominally $sp^2$
C site  leads to some distortion
of the original ring structure. The improvement of the
spectrum is questionable.
Hence, it is clear that we need to look for a
better choice than boron. In fact, Fig.~4 suggests that the Al(AsQ)$_3$
provides an excellent improvement over the more usual AlQ$_3$ while
its synthesis would not be as difficult as the boron substituted
form.

\section{conclusion}

We show that the spectra of metal chelated 8-hydorxyquinolines can be
manipulated and improved to obtain much better luminescence
properties.
 We have shown how the spectrum can be shifted and
that the intensity would increase if the triplet
excitations could be harvested. The replacement of the
N in AlQ$_3$ by an As atom provides a definite improvement,
while the further substitution of the {\it p-}Carbon by B is
not seen to provide a useful improvement.
This is a preliminary study which anticipates
a more detailed calculation where the $L-S$ coupling would be
explicitly included.

\appendix*
\section{
Some considerations of the spin-orbit interaction}
The relativistic effects for heavy atoms can be included up to
$\alpha^2$, where $\alpha$ is the fine structure constant, in
first principles pseudopotentials. The result splits into
two terms of the form:
\begin{equation}
V_{rel}(r)=\sum_l |l>[V_l(r)+V_l^{so}(r){\bf L.S}]<l|
\end{equation}
where the  $l$-sum  is a sum over angular momentum states.
 The first term is the
scalar part, and is not significant for the problem of the
singlet-triplet bottleneck. The second term in the above equation,
$V_2=\sum_l |l>V_l^{so}(r){\bf L.S}<l|$
introduces mixing between orbital and spin angular momentum states,
leading to the break down of the
 usual spin selection rules. However, the full inclusion
of the spin-orbit term in first-principles CI calculations of the
type discussed in this paper is numerically quite prohibitive, given
the large number of electrons which need to be handled
 in typical LED materials. Leave aside CI calculations, the
numerical evaluation even with a pseudopotential scheme is 
extremely demanding.\cite{joanno}

The spin-orbit operator is highly local in the sense that it is mostly
sensitive to changes in electron density close to the nucleus, where
relativistic effects dominate. Thus we may look at calculations for
systems containing As as a guide.
In order to get at least a simple ``grosso modo'' estimate,
 we may compare the homo-lumo
gap (bandgap) of As obtained from a non-relativistic (i.e, standard)
calculation with a calculation where some effort is put into
an evaluation of the relativistic effects.\cite{gonze} Thus Gonze et al.
have carried out calculations which can be used for a simple estimate
of the effect of the spin-orbit effect in As. The s-levels in As are
about 9 eV lower than the p-levels. The p-levels suffer a splitting of
about 0.36 eV in As, and is a measure of the strength of the spin-orbit
 term in As. Since we are concerned with singlet and triplet excitons,
we should compare this spin-orbit splitting with the gap 
 between hole energies and the Fermi energy, and electron energies and
the Fermi energy in As. Numerical values for these, for As have been given by
J.-P. Issi,\cite{issi}, and by Priestly et al.,\cite{priestly}. 
Issi gives 0.154 eV and 0.202 eV for electrons and holes, while
Priestly et al give 0.177 and 0.190 eV. Hence the total effect
for an electron-hole pair is of the order of 0.35 eV by one estimate, and 
0.36 eV by the second estimate. These numbers are in fact very similar to the
spin-orbit energy ($\sim$ 0. 36 eV).
These considerations suggest that when estimating the oscillator
strengths for the processes $<eh(\uparrow,\uparrow)|\vec{r}|$groundstate$>$
and  $<eh(\uparrow,\downarrow)|\vec{r}|$groundstate$>$, we may, as a
{\it grosso  modo} estimate take the oscillator strengths to be 
of the same magnitude.


\begin{table*}
\caption
{
 Transition Energies $E$ in  eV, 
Oscillator Strengths (a.u.), $f_{os}$,
 for AlQ$_3$, Al(AsQ)$_3$, and Al(AsBQ)$_3$.
Transition type is labeled S or T to indicate singlet or triplet character 
for no L-S coupling ( thus, e.g.,  T7 is the 7$^{th}$ triplet).
The oscillator strength $f_o$ for triplets are not given.
The transitions marked with an asterisk are the
ones which are most closely identifiable with a HOMO $\to$ LUMO
transition. This  identification is not very satisfactory in the 
Al(AsBQ)$_3$ molecule (see the text)
}
\begin{ruledtabular}
\begin{tabular}{cccccccccccc}
 Transition    & AlQ$_3$& &Transition     &Al(AsQ)$_3$& &Transition &Al(AsBQ)$_3$& $\,$  & \\       
 type& $E$      &$f_{os}$ & type  & $E$ &      $f_{os}$  &  type  & $E$         &$f_{os}$   & \\
                                                                                 \hline \\
 T1 &2.152   &       &  T1 &1.847   &       &  T1   &1.870   &          & \\
 T2 &2.190   &       &  T2 &1.854   &       &  T2   &1.932   &          &  \\
 T3 &2.228   &       &  T3 &1.867   &       &  T3   &1.945   &          &  \\ 
 S1*&2.718   &0.0047&  T4* &2.613   &       &  T4   &2.102   &          & \\
 T4*&2.741   &      &  S1* &2.631   &0.0129 &  T5   &2.105   &          &  \\
 S2 &2.909   &0.0590 &  T5 &2.744   &       &  T6   &2.181   &          &\\
 S3 &2.924   &0.0180 &  S2 &2.769   &0.0060 &  S1   &2.333   &0.0008    &    \\
 T5 &2.925   &       &  S3 &2.897   &0.1066 &  S2   &2.337   &0.0015    & \\
 S4 &3.000   &0.0423 &  S4 &2.998   &0.0265 &  S3   &2.409   &0.0011    &      \\
 S5 &3.029   &0.0027 &  T6 &3.002   &       &  T7   &2.595   &          &    \\
 T6 &3.056   &       &  T7 &3.054   &       & *T8   &2.609   &          &   \\
 S6 &3.187   &0.0173 &  S5 &3.116   &0.0242 & *S4   &2.614   &0.0169    & \\
 T7 &3.228   &       &  S6 &3.195   &0.0244 &  S5   &2.643   &0.0138    & \\
 T8 &3.255   &       &  T8 &3.232   &       &  T9   &2.753   &          &    \\
 S7 &3.273   &0.0049 &  S7 &3.324   &0.0741 &  S6   &2.756   &0.0004    &    \\
\end{tabular}   
\end{ruledtabular}                                                        
\label{table1}                                                          
\end{table*}



\vspace{0.2in}

Fig.1 Structure of the light emitter AlQ$_3$.  See ref.\cite{structure}  for
      bond lengths, angles etc.\\

Fig.2 N atoms of the quinolines have been replaced by As in Al(AsQ)$_3$ \\

Fig.3 The C atom in the {\it para}-position to N has been replaced by
     B in  Al(AsBQ)$_3$. The Kekule-type structure is only schematic. \\

Fig.4 Calculated spectra of the three compounds. Solid curves: only the
      singlet is available for light emission; dashed curves:
      the triplet exciton is also  harvested (using a $f_{os}$ comparable
      to the nearest singlet (see appendix).
	 Note that the intensity axis is logarithmic.
      The calculation
      is for isolated molecules, and hence a broadening of 0.025 eV has been
      included.

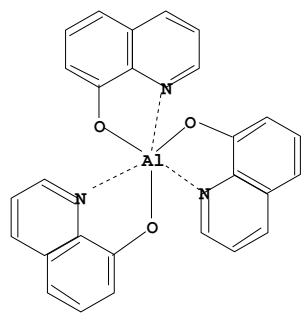
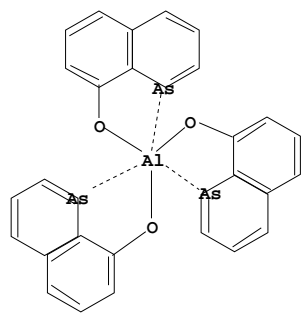
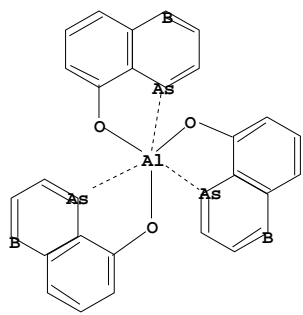
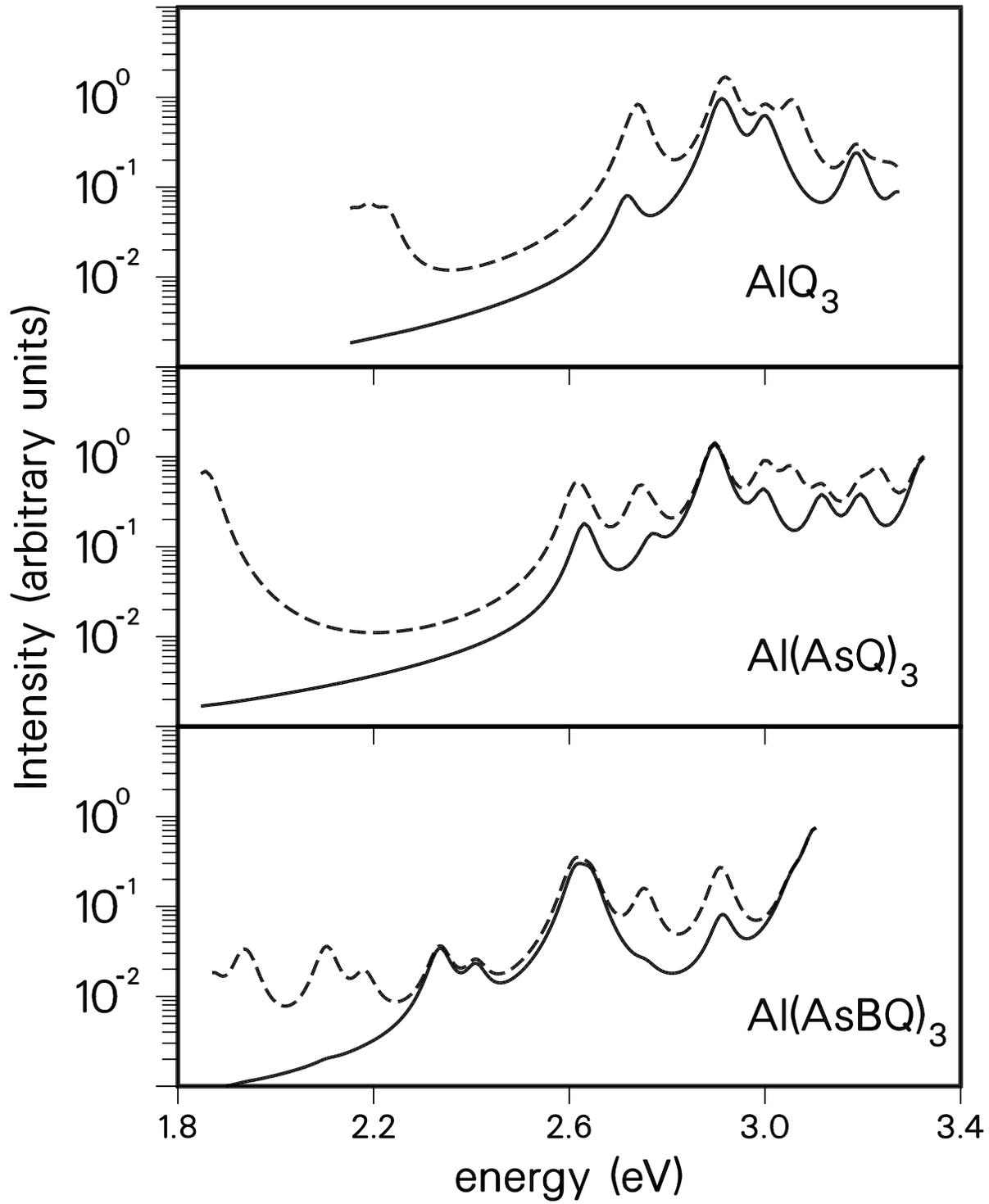
\end{document}